# Field normalization of scientometric indicators[*]

Ludo Waltman and Nees Jan van Eck

Centre for Science and Technology Studies, Leiden University, The Netherlands
{waltmanlr, ecknjpvan}@cwts.leidenuniv.nl

When scientometric indicators are used to compare research units active in different scientific fields, there often is a need to make corrections for differences between fields, for instance differences in publication, collaboration, and citation practices. Field-normalized indicators aim to make such corrections. The design of these indicators is a significant challenge. We discuss the main issues in the design of field-normalized indicators, and we present an overview of different approaches that have been developed for dealing with the problem of field normalization. We also discuss how field-normalized indicators can be evaluated, and we consider the sensitivity of scientometric analyses to the choice of a field normalization approach.

## 1. Introduction

Many scientometric analyses are restricted to a single field of science, but scientometric analyses also commonly stretch out over multiple scientific fields, and they often even aim to cover the entire scientific universe. University rankings for instance rely on scientometric indicators that are supposed to provide meaningful information about the performance of universities across many different fields of science. Likewise, many universities regularly carry out scientometric analyses in which they compare their performance in different scientific fields.

Scientific fields of course differ from each other in many ways, and some of these differences have important implications for scientometric analyses. For instance, in some fields researchers tend to produce many more outputs than in other fields. In some fields researchers focus on publishing journal articles, while in other fields they are more interested in writing books. In some fields researchers work together in large collaborative teams, often resulting in publications with many co-authors, while in other fields researchers prefer to work individually or in small teams. In some fields researchers cite a lot, while in other fields they cite much more sparingly, and in some fields they mainly cite recent work, while in other fields they prefer to cite older literature.

---

[*] Forthcoming in Glänzel, W., Moed, H.F., Schmoch U., & Thelwall, M. (2018). *Springer Handbook of Science and Technology Indicators*. Springer.

Given these differences between scientific fields, it is clear that the interpretation of a scientometric analysis that covers multiple fields is far from straightforward. Suppose that a biologist has produced 25 publications during the past five years, while an economist has produced 10 publications during the same time period. Can it be concluded that the biologist has been more productive than the economist? This depends on our understanding of the concept of productivity. If productivity is understood simply as the number of publications produced during a certain time period, the biologist obviously has been more productive than the economist. However, in many cases we are probably interested in a more sophisticated concept of productivity. We may have in mind a concept of productivity that accounts for differences between fields in the rate at which researchers tend to produce publications. Based on such a more refined notion of productivity, the answer to our question is much less obvious. It may actually turn out that from this perspective the economist should be considered more productive than the biologist.

To capture the more sophisticated concept of productivity suggested above, we need a scientometric indicator that in some way corrects for differences between scientific fields in the typical number of publications produced by a researcher. Such an indicator is referred to as a field-normalized indicator. Field-normalized indicators can be constructed not only for the concept of productivity, but also for other scientometric concepts. In the literature, field-normalized indicators of scientific impact, calculated based on citation counts, have received most attention, and they will also play a prominent role in this chapter.

The design of field-normalized indicators is a significant challenge. In this chapter, we discuss the main issues in the design of these indicators. We present an overview of different approaches that have been developed for dealing with the problem of field normalization. We also discuss how field-normalized indicators can be evaluated, and we consider the sensitivity of scientometric analyses to the choice of a field normalization approach.

This chapter partly builds on a recent review of the literature on citation impact indicators published by one of us (Waltman, 2016a).

## 2. What is field normalization?

It is notoriously difficult to define in a precise way what is meant by field normalization. A precise definition of the idea of field normalization requires a definition of the notion of a field. It also requires a clear perspective on the way in which scientometric indicators are affected by differences between fields. As we will explain below, these requirements are challenging, and therefore the idea of field normalization will almost inevitably remain somewhat ill-defined.

Defining the notion of a field is far from straightforward. There is a lack of standardized terminology. No agreement exists on the differences between the term 'field' and terms such as 'area', 'discipline', 'domain', 'specialty', and 'topic'. In fact, these terms often seem to be used more or less interchangeably. More fundamentally, the idea of a field can be conceptualized in different ways. A useful overview of different conceptualizations is provided by Sugimoto and Weingart (2015). They distinguish between cognitive, social, communicative, and institutional perspectives as well as perspectives based on separatedness and tradition. Each of these perspectives provides a different understanding of the idea of a field.

Defining the notion of a field is made even more difficult by the fact that science is structured in a hierarchical way, allowing fields to be identified at different hierarchical levels (Zitt, Ramanana-Rahary, & Bassecoulard, 2005). For instance, depending on the hierarchical level that one prefers, citation analysis, bibliometrics, information science, and social sciences could all be seen as fields. Moreover, even when one focuses as much as possible on a single hierarchical level, fields typically will not be neatly separated from each other. For instance, bibliometrics, scientometrics, and research evaluation perhaps could be regarded as fields at more or less the same hierarchical level. However, it is clear that these fields are strongly interrelated and have a considerable overlap.

Field normalization of scientometric indicators is motivated by the idea that differences between fields lead to distortions in scientometric indicators. One could think of this in terms of signal and noise. Scientometric indicators provide a signal of concepts such as productivity or scientific impact, but they are also affected by noise. This noise may partly be due to differences between fields, for instance differences in publication, collaboration, and citation practices. Field normalization aims to remove this noise while maintaining the signal.

However, the distinction between signal and noise is much less clear than it may seem at first sight. To illustrate this, let's consider citation-based indicators of scientific impact. Publications in information science on average are cited much less frequently than publications in for instance the life sciences. A citation-based indicator that does not account for this may be considered very noisy. The indicator may be seen as strongly biased against information science research. Suppose therefore that we use an indicator that corrects for differences in 'citation density' between information science and other fields. Let's now zoom in on information science. Within information science, publications in scientometrics on average receive significantly more citations than publications in for instance library science. Again, we may feel that our indicator is too noisy and that we need to get rid of the noise. Consequently, suppose that we use an indicator that corrects for differences in citation density not only between information science and other fields, but also between scientometrics and other subfields within information science. We now zoom in on scientometrics. Within scientometrics, publications on citation analysis tend to receive more citations than publications on a topic such as co-authorship analysis. This may also be seen as noise that we need to get rid of. The next step then may be to use an indicator that corrects for differences in citation density even between different topics within scientometrics. However, we could of course argue that even this indicator is noisy. Suppose that empirical publications on citation analysis are cited more frequently than theoretical publications. This could then be claimed to show that the empirical publications have a higher impact than the theoretical publications. On the other hand, we could also argue that empirical and theoretical research on citation analysis represent two different subtopics and that we need to get rid of noise that is due to differences in citation density between these subtopics. However, if we keep following such a reasoning, at some point everything is considered noise and there is no signal left, meaning that indicators become completely non-informative.

The above example illustrates that there is no objective way of distinguishing between signal and noise. We may say that scientometric indicators are distorted by noise that is due to differences between fields. However, fields can reasonably be defined at different hierarchical levels, leading to different perspectives on what should count as signal and what should be seen as noise. When working with field-normalized indicators, choosing a certain hierarchical level for defining fields, and consequently making a certain distinction between signal and noise, is a normative decision. There is no objective way in which this choice can be made. Probably there is agreement that fields should not be defined in a very broad or very narrow way, but this still leaves open many intermediate ways in which fields can be defined. A single optimal way of defining fields does not exist (Zitt et al., 2005). Ideally, the hierarchical level at which fields are defined is chosen in such a way that it aligns well with the purpose of a specific scientometric analysis. In some analyses it may be desirable to work with relatively narrow fields, while in other analyses broader fields may be appropriate.

## 3. Field classification systems

Most field-normalized indicators require an operationalization of scientific fields. We refer to such an operationalization as a field classification system. Different types of field classification systems can be distinguished. We make a distinction between classification systems of journals, publications, and researchers. Many different classification systems exist. We do not aim to provide a comprehensive overview of these systems. Instead, we focus specifically on classification systems that have been used for field normalization purposes, either in the scientometric literature or in applied scientometric work. Each of the classification systems discussed below deals in a different way with the challenges in operationalizing scientific fields.

### 3.1. Field classification systems of journals

The field classification systems used most frequently by field-normalized indicators are journal-based systems. In these systems, each journal is assigned to one or more fields. Some journal-based classification systems do not allow fields to overlap. A journal can be assigned to only one field in such systems. However, in most journal-based classification systems, overlap of fields is allowed, in which case a journal may belong to multiple fields. Some journal-based classification systems have a hierarchical structure and consist of multiple levels. Each field at a lower level is then considered to be part of a field at a higher level.

The Web of Science (WoS) database offers a classification system in which each journal indexed in the database is assigned to one or more fields. These fields are referred to as categories. There are about 250 fields in the WoS classification system.

A somewhat similar classification system is made available in the Scopus database. This system is referred to as the All Science Journal Classification (ASJC). The system has a hierarchical structure consisting of two levels. There are over 300 fields at the bottom level. These fields have been aggregated into 27 fields at the top level. Each journal indexed in Scopus belongs to one or more fields. A comparison of the accuracy of the WoS and Scopus classification systems is reported in a study by Wang and Waltman (2016). According to this study, the WoS classification system is significantly more accurate than the Scopus classification system.

In the Essential Science Indicators, a tool that is based on the WoS database, a classification system of 22 broad fields is made available. In this system, it is not possible for a journal to belong to multiple fields. Each journal is assigned exclusively to a single field.

Other journal-based classification systems include the classification system of the US National Science Foundation, the classification system developed by Science-Metrix, and the classification system of Glänzel and Schubert (2003). The classification system of the National Science Foundation covers 125 fields, which have been aggregated into 13 broad fields. A journal can belong to only one field in this system. For a long time, the system has been used in the Science and Engineering Indicators reports prepared by the National Science Foundation. Science-Metrix is a company specialized in research evaluation that has developed its own classification system. This system is made freely available. It includes 176 fields, aggregated into 22 broad fields, with each journal being assigned to only one field. We refer to Archambault, Beauchesne, and Caruso (2011) for more details on the approach that was taken to construct the Science-Metrix classification system. The classification system of Glänzel and Schubert (2003) consists of two levels. The 67 fields at the bottom level have been aggregated into 15 fields at the top level. Journals may be assigned to more than one field in this system.

Multidisciplinary journals with a broad scope represent a significant challenge for journal-based classification systems. *Nature*, *PNAS*, and *Science* are well-known examples of such journals. Other examples are open access 'mega journals' such as *PLOS ONE* and *Scientific Reports*. In a journal-based classification system, these multidisciplinary journals are typically assigned to a special category. In the WoS categories classification system, this category is for instance called *Multidisciplinary Sciences*. In the Scopus ASJC classification system, it is referred to as

*Multidisciplinary.* The use of a special category for multidisciplinary journals is problematic because such a category clearly does not represent a scientific field. In practice, this problem is often addressed by creating a publication-based classification system for publications in multidisciplinary journals and by complementing a journal-based classification system with such a publication-based classification system. This approach, introduced by Glänzel, Schubert, and Czerwon (1999; see also Glänzel & Schubert, 2003), has been widely adopted. Of course, there always is some arbitrariness in deciding which journals should be considered multidisciplinary. It is clear that journals such as the ones mentioned above are of a multidisciplinary nature. However, it may be argued that journals such as *The Lancet*, *New England Journal of Medicine*, and *Physical Review Letters* should also be considered multidisciplinary and that it would be preferable to create a publication-based classification system for publications in these journals.

### 3.2. Field classification systems of publications

Instead of journal-based field classification systems, it is also possible to use publication-based field classification systems. Publication-based classification systems potentially offer a more accurate and more fine-grained representation of scientific fields than their journal-based counterparts. Most publication-based classification systems are restricted to a single scientific discipline. Algorithmically constructed classification systems are an exception and may cover all scientific fields.

There are various scientific disciplines that have their own publication-based classification system. These systems often have a hierarchical structure, and they usually allow publications to be assigned to multiple fields. The use of these systems in field-normalized indicators has been studied by Bornmann, Mutz, Neuhaus, and Daniel (2008), Neuhaus and Daniel (2009), Radicchi and Castellano (2011), and Van Leeuwen and Calero-Medina (2012). These authors have focused on respectively the Medical Subject Headings, the Chemical Abstracts sections, the Physics and Astronomy Classification Scheme, and the EconLit classification system. Like in the case of the journal-based classification systems discussed in Subsection 3.1, it is important to be aware that publication-based classification systems such as the ones mentioned above have not been designed specifically for field normalization purposes.

Publication-based classification systems that are constructed algorithmically may cover all scientific fields rather than only fields within a single discipline. An approach for the algorithmic construction of publication-based classification systems has been proposed by Waltman and Van Eck (2012). In this approach, a classification system is constructed by clustering publications based on direct citation relations. Each publication is assigned to only one field. The use of algorithmically constructed publication-based classification systems in field-normalized indicators has been studied by Ruiz-Castillo and Waltman (2015) and Perianes-Rodriguez and Ruiz-Castillo (2017). A practical application can be found in the CWTS Leiden Ranking, a bibliometric ranking of major universities worldwide that is available at www.leidenranking.com. In this ranking, citation-based indicators of scientific impact are normalized using an algorithmically constructed publication-based classification system in which about 4000 scientific fields are distinguished.

An algorithmic approach to the construction of a publication-based classification system is also taken in Microsoft Academic, a recently introduced bibliometric data source somewhat similar to Google Scholar. Hug, Ochsner, and Brändle (2017) have found that fields in the classification system of Microsoft Academic are too specific and not coherent, leading them to conclude that the classification system of Microsoft Academic is not suitable for field normalization purposes.

**3.3. Field classification systems of researchers**

Field classification systems of researchers represent a quite different approach to operationalize scientific fields. The use of researcher-based classification systems in field-normalized indicators is much less common than the use of journal-based and publication-based classification systems. Below we discuss two researcher-based classification systems that have been used in the scientometric literature.

Giovanni Abramo and Ciriaco Andrea D'Angelo have published a large number of papers in which they use the official classification system of Italian researchers (e.g., Abramo & D'Angelo, 2014). This is a hierarchical system consisting of two levels. At the top level, 14 fields are distinguished. These fields are referred to as university disciplinary areas. At the bottom level, there are 370 fields, referred to as specific disciplinary sectors, with each specific disciplinary sector being part of a single university disciplinary area. In Italy, each researcher at a university must belong to exactly one specific disciplinary sector. We will get back to the Italian classification system of researchers in Section 5.

Another example of a classification system of researchers is the classification system of the Mendeley reference management tool. This system has been used by Bornmann and Haunschild (2016a) in a proposal for a field-normalized indicator of scientific impact based on Mendeley reader counts. In the Mendeley classification system, a distinction between 28 fields is made. Each Mendeley user is able to assign him- or herself to one of these 28 fields.

## 4. Overview of field-normalized indicators

In this section, we provide an overview of field-normalized indicators that have been proposed in the scientometric literature. The literature on field-normalized indicators is extensive. We therefore do not discuss each individual proposal presented in the literature. Instead, our focus is on what we consider to be the more significant contributions that have been made. Other contributions may not be covered or may be mentioned only very briefly. We also do not aim to give a historical overview of the literature. We discuss important ideas that have been presented in the literature, but we do not necessarily trace the historical development of these ideas.

In principle, field-normalized variants can be developed for any type of scientometric indicator. In practice, however, scientometricians have put most effort into the development of field-normalized indicators of the impact of scientific publications, where impact is typically operationalized using citations. Our focus in this section is therefore mostly on field-normalized indicators of impact, although we also discuss field-normalized indicators of productivity. Most of the indicators that we consider in this section rely on field classification systems such as the ones introduced in the previous section, but we also discuss indicators that do not require a field classification system.

Field-normalized indicators typically normalize not only for the field of a publication, but also for the age of a publication. This is important in the case of indicators based on citations, since older publications have had more time to receive citations than younger publications. Indicators may also normalize for other characteristics of a publication. For instance, they sometimes normalize for publication type, where a distinction can be made between categories such as research article, review article, and letter.

**4.1. Indicators of impact: Indicators based on normalized citation scores**

The normalized citation score of a publication can be defined in different ways. The most straightforward approach is to define it as the ratio of the actual and the expected number of citations of a publication, where the expected number of citations of a publication equals the average number of citations of all publications in the same field and in the same publication year (and often also in the same publication type category). Whether publications are in the same field is determined based on a field classification system, such as one of the systems discussed in Section 3.

In order to obtain indicators at the level of for instance a research group, a research institution, or a journal, the normalized citation scores of individual publications need to be aggregated. This is typically done either by averaging or by summing the normalized citation scores. Averaging the scores yields a so-called size-independent indicator of impact, while summing the scores gives a size-dependent impact indicator. These indicators are known under various different names. The size-independent indicator is for instance known as the mean normalized citation score (Waltman, Van Eck, Van Leeuwen, Visser, & Van Raan, 2011a), the item-oriented field normalized citation score average (Lundberg, 2007), the category normalized citation impact (in the commercial InCites tool), and the field weighted citation impact (in the commercial Scopus and SciVal tools). The size-dependent indicator is sometimes referred to as the total normalized citation score (Waltman et al., 2011a).

A recent development is the application of the above approach for calculating field-normalized impact indicators to bookmarks in Mendeley instead of citations. Studies of field-normalized indicators based on Mendeley bookmarks, often interpreted in terms of readership, have been reported by Fairclough and Thelwall (2015a) and Haunschild and Bornmann (2016).

A number of alternative approaches have been explored for defining the normalized citation score of a publication. One alternative is to leave out non-cited publications from the calculation of the expected number of citations of a publication (Abramo, Cicero, & D'Angelo, 2012a, 2012b). Another alternative is to determine the expected number of citations of a publication based on the idea of so-called exchange rates, where the similarity between fields in the shape of citation distributions is used to determine how many citations in one field can be considered equivalent to a given number of citations in another field (Crespo, Herranz, Li, & Ruiz-Castillo, 2014;

Crespo, Li, & Ruiz-Castillo, 2013). A third alternative is to apply a logarithmic transformation to the citation counts of publications (Fairclough & Thelwall, 2015b; Lundberg, 2007; Thelwall & Sud, 2016; Thelwall, 2017). A fourth alternative is to transform citation counts into *z*-scores (Bornmann & Daniel, 2009; Vaccario, Medo, Wider, & Mariani, 2017; Zhang, Cheng, & Liu, 2014). This approach can be combined with a logarithmic transformation of citation counts (Lundberg, 2007). A fifth alternative is to transform citation counts using a two-parameter power-law function (Radicchi & Castellano, 2012c). Finally, a sixth alternative proposed in the literature is to transform citation counts into binary variables based on whether publications have been cited or not (Thelwall, 2017).

There has also been considerable discussion in the literature about the best way to calculate field-normalized impact indicators at aggregate levels, for instance at the level of research groups or research institutions (e.g., Gingras & Larivière, 2011; Lundberg, 2007; Moed, 2010b; Opthof & Leydesdorff, 2010; Van Raan, Van Leeuwen, Visser, Van Eck, & Waltman, 2010; Vinkler, 2012; Waltman et al., 2011a). The approach discussed above, in which normalized citation scores of individual publications are averaged or summed, nowadays is the most commonly used approach. An alternative approach is to calculate the average or the sum of the actual citation counts of a set of publications and to divide the outcome by the average of the expected citation counts of the same set of publications (e.g., Glänzel, Thijs, Schubert, & Debackere, 2009; Moed, De Bruin, & Van Leeuwen, 1995; Van Raan, 2005). In this alternative approach, normalization can be considered to take place at the level of an oeuvre of publications rather than at the level of individual publications (Moed, 2010b). When an analysis includes publications from multiple fields or multiple years, normalization at the oeuvre level will generally yield results that are different from the outcomes obtained by normalizing at the level of individual publications. We refer to Larivière and Gingras (2011), Waltman, Van Eck, Van Leeuwen, Visser, and Van Raan (2011b), and Herranz and Ruiz-Castillo (2012) for empirical analyses of the differences between the two approaches.

Another issue in the calculation of field-normalized impact indicators at aggregate levels is the choice of a counting method for handling co-authored publications. Full and fractional counting are the two most commonly used counting methods. In the case of full counting, each publication is fully counted for each co-author. On the other hand, in the case of fractional counting, a publication with *n* co-authors is counted with a weight of 1 / *n* for each co-author. The choice of a counting method influences the extent to which an indicator can be considered to provide properly field-normalized statistics (Waltman & Van Eck, 2015). We will get back to this issue in Subsection 5.1.

**4.2. Indicators of impact: Indicators based on percentiles**

Percentile-based impact indicators value publications based on their position in the citation distribution of their field and publication year, where fields are defined using a field classification system, for instance one of the systems discussed in Section 3. In the most straightforward case, these indicators make a distinction between lowly and highly cited publications. For instance, all publications that in terms of citations belong to the top 10%, top 5%, or top 1% of their field and publication year may be regarded as highly cited, as suggested by Tijssen, Visser, and Van Leeuwen (2002) and Van Leeuwen, Visser, Moed, Nederhof, and Van Raan (2003). A generalization of this idea has been proposed by Leydesdorff, Bornmann, Mutz, & Opthof (2011). In their proposal, a number of classes of publications are distinguished. Each class of publications is defined in terms of percentiles of the citation distribution of a field and publication year. The first class may for instance include all publications whose number of citations is below the 50th percentile of the citation distribution of their field and publication year, the second class may include all publications whose number of citations is between the 50th and the 75th percentile, and so on. In the proposed approach, publications are valued based on the class to which they belong. Publications in the lowest class have a value of one, publications in the second-lowest class have a value of two, and so on.

A difficulty in the calculation of percentile-based indicators is the issue of ties, that is, multiple publications with the same number of citations. Suppose we want to identify the 10% most frequently cited publications in a certain field and publication year. We then need to find a threshold such that exactly 10% of the publications in this field and publication year have a number of citations that is above the threshold. In practice, it usually will not be possible to find such a threshold. Because of the issue of ties, typically any threshold will yield either too many or too few publications whose number of citations is above the threshold. This means that fields cannot be made fully comparable, since the distortion caused by the issue of ties will be different in different fields. In the literature, various approaches for dealing with the issue of ties have been explored (e.g., Leydesdorff et al., 2011; Pudovkin & Garfield, 2009; Van Leeuwen et al., 2003; Waltman & Schreiber, 2013). We refer to Waltman and Schreiber (2013) for a summary of these approaches and to Schreiber (2013) for an empirical comparison.

Field-normalized impact indicators can also be constructed by combining the idea of percentile-based indicators with the idea of indicators based on normalized citation scores. Such an approach has been introduced by Albarrán, Ortuño, & Ruiz-Castillo (2011a, 2011b). In the proposed approach, indicators are used to characterize the distribution of citations over the highly cited publications in a field. The indicators resemble indicators developed in the field of economics for characterizing income distributions.

Glänzel (2013) and Glänzel, Thijs, and Debackere (2014) have proposed indicators that, like the above-mentioned indicators proposed by Leydesdorff et al. (2011), distinguish between a number of classes of publications. However, instead of percentiles, these indicators rely on the method of characteristic scores and scales (Glänzel & Schubert, 1988) to define the classes. Publications belong to the lowest class if they have fewer citations than the average of their field, they belong to the second-lowest class if they do not belong to the lowest class and if they have fewer citations than the average of all publications not belonging to the lowest class, and so on. An alternative approach is to define the classes based on median instead of average citation counts (Vîiu, 2017).

Percentile-based approaches may also be used to normalize altmetric indicators. Bornmann and Haunschild (2016c) have suggested a percentile-based approach for normalizing Twitter counts.

**4.3. Indicators of impact: Indicators that do not use a field classification system**

All field-normalized indicators discussed so far rely on a field classification system that operationalizes scientific fields. As discussed in Sections 2 and 3, the operationalization of fields is a difficult problem. Field classification systems offer a simplified representation of fields. By necessity, any field classification system relies partly on arbitrary and contestable choices. In this subsection, we discuss field-normalized impact indicators with the attractive property that they do not require a field classification system.

An approach that has been explored in the literature is to identify for each publication a set of similar publications, allowing the citation score of the focal publication to be compared with the citation scores of the identified similar publications. Similar publications may be identified based on shared references (i.e., bibliographic coupling relations), as suggested by Schubert and Braun (1993, 1996). Alternatively, as demonstrated by Colliander (2015), the identification of similar publications may be done based on a combination of shared references and shared terms. Another possibility is to use co-citation relations to identify similar publications. This idea is used in the relative citation ratio indicator, an indicator introduced by a research team at the US National Institutes of Health (Hutchins, Yuan, Anderson, & Santangelo, 2016) that has attracted a significant amount of attention. We refer to Janssens, Goodman, Powell, and Gwinn (2017) for a critical discussion of the relative citation ratio indicator (for a response by the original authors, see Hutchins, Hoppe, Meseroll, Anderson, & Santangelo, 2017). Instead of working at the level of individual publications, it is also possible to work at the journal level. The citation score of a journal can then be compared with the citation scores of other similar journals. The latter journals may be identified based on citations given to the focal journal (Dorta-González, Dorta-González, Santos-Peñate, & Suárez-Vega, 2014).

Another field normalization approach that does not require a field classification system is known as citing-side normalization (Zitt & Small, 2008), sometimes also referred to as fractional citation weighting (Zitt & Small, 2008), fractional citation counting (Leydesdorff & Opthof, 2010), source normalization (Moed, 2010a), or a priori normalization (Glänzel, Schubert, Thijs, & Debackere, 2011). Citing-side normalization is based on the idea that differences between fields in citation density are to a large extent caused by the fact that in some fields publications tend to have longer reference lists than in other fields. Citing-side normalization performs a correction for the length of the reference list of citing publications. The basic idea of citing-side normalization can be implemented in different ways. One possibility is to correct for the average reference list length of citing journals (Zitt, 2010; Zitt & Small, 2008). Another possibility is to correct for the reference list length of individual citing publications (Glänzel et al., 2011; Leydesdorff & Bornmann, 2011; Leydesdorff & Opthof, 2010; Leydesdorff, Zhou, & Bornmann, 2013; Moed, 2010a). A combination of these two options is possible as well, and this is how citing-side normalization is implemented in the current version of the source normalized impact per paper (SNIP) journal impact indicator (Waltman, Van Eck, Van Leeuwen, & Visser, 2013).

Instead of correcting for reference list length on the citing side, an alternative approach is to correct for reference list length on the cited side. In this approach, a correction can be made for either the reference list length of a cited publication (Kosmulski, 2011) or the average reference list length of a cited journal (Franceschini, Galetto, Maisano, & Mastrogiacomo, 2012; Franceschini & Maisano, 2014; Nicolaisen & Frandsen, 2008). A third possibility is to correct for the average reference list length of all publications belonging to the same field as a cited publication (Bornmann & Haunschild, 2016b). However, this again requires a field classification system, just like in the case of the indicators discussed in Subsections 4.1 and 4.2.

Recursive impact indicators, first introduced by Pinski and Narin (1976) and often inspired by the well-known PageRank algorithm (Brin & Page, 1998), offer another approach that is related to the idea of citing-side normalization. Examples of recursive impact indicators are the eigenfactor and article influence indicators of journal impact (Bergstrom, 2007; West, Bergstrom, & Bergstrom, 2010) and the SCImago journal rank (SJR) indicator (González-Pereira, Guerrero-Bote, & Moya-Anegón, 2010; Guerrero-Bote & Moya-Anegón, 2012). We refer to Waltman and Yan (2014) and Fragkiadaki and Evangelidis (2014) for overviews of the literature on recursive impact indicators and to Waltman and Van Eck (2010) for a discussion of the relation between these indicators and citing-side normalized indicators.

**4.4. Indicators of productivity**

Although field-normalized indicators of impact have received most attention in the scientometric literature, some attention has also been given to field-normalized indicators of productivity (sometimes also referred to as efficiency). Productivity indicators can for instance be calculated for researchers, research groups, and research institutions. A simple productivity indicator is the average number of publications produced per researcher. A more advanced productivity indicator may also take into account the number of citations publications have received. Field-normalized productivity indicators perform a correction for differences between fields in the rate at which publications are produced and citations are received.

Field-normalized productivity indicators play a prominent role in the work of Giovanni Abramo and Ciriaco Andrea D'Angelo. In particular, Abramo and D'Angelo make extensive use of an indicator referred to as the fractional scientific strength (FSS). For an individual researcher, FSS essentially equals the sum of the normalized citation scores (see Subsection 4.1) of the publications of the researcher divided by the salary of the researcher. Likewise, for a group of researchers working in the same field, FSS equals the sum of the normalized citation scores of their publications divided by their total salary. When FSS is calculated for a group of researchers working in different fields, for instance all researchers affiliated with a particular research institution, a correction needs to be made for differences between fields in the average publication output and the average salary of researchers. One way in which this can be done is by first calculating each researcher's field-normalized FSS, defined as the researcher's FSS divided by the average FSS of all researchers working in the same field, and by then calculating the average field-normalized FSS of all researchers. We refer to Abramo and D'Angelo (2014) for a more detailed discussion of the calculation of the FSS indicator. For a discussion of an alternative productivity indicator, based on highly cited publications instead of normalized citation scores, we refer to Abramo and D'Angelo (2015b).

In practice, calculating the FSS indicator is highly challenging because it requires data on the publications and the salaries of all researchers working in a field. Abramo and D'Angelo address this difficulty by taking into account only Italian publications and Italian researchers in the calculation of the FSS indicator. In Italy, unlike most other countries, the data required for the calculation of the FSS indicator is available. Abramo and D'Angelo calculate normalized citation scores of publications using the WoS journal-based field classification system (see Subsection 3.1). However, they also need a second classification system. To calculate researchers' field-normalized FSS, they rely on a classification system of Italian researchers (see Subsection 3.3).

In most countries, the data needed to calculate field-normalized productivity indicators is not available. Obtaining productivity indicators that allow for meaningful cross-country comparisons is even more challenging, as pointed out by Aksnes, Sivertsen, Van Leeuwen, and Wendt (2017). An interesting proposal for calculating field-normalized productivity indicators, even when only limited data is available, has been presented by Koski, Sandström, and Sandström (2016). This proposal focuses on the difficulty of researchers that have no publications in a certain time period. These researchers are invisible in databases such as WoS and Scopus, which causes problems when using these databases to calculate productivity indicators. To deal with this issue, a statistical methodology is proposed for estimating the number of researchers without publications.

A number of studies have focused specifically on designing field-normalized indicators of the productivity of individual researchers. In particular, several proposals have been made for variants of the *h*-index (Hirsch, 2005) that correct for field differences (Batista, Campiteli, Kinouchi, & Martinez, 2006; Harzing, Alakangas, & Adams, 2014; Iglesias & Pecharromán, 2007; Kaur, Radicchi, & Menczer, 2013; Radicchi, Fortunato, & Castellano, 2008). Other interesting proposals for comparing individual researchers active in different fields have been presented by Kaur, Ferrara, Menczer, Flammini, and Radicchi (2015) and Ruocco and Daraio (2013).

It is important to be aware of the difference between productivity indicators and size-independent impact indicators. Both types of indicators are independent of size, which is convenient for instance when making comparisons between larger and smaller research institutions. However, the two types of indicators are based on fundamentally different notions of size. Productivity indicators take an input perspective on the notion of size, for instance the number of researchers affiliated with an institution. Size-independent impact indicators take an output perspective on the notion of size, namely the number of publications produced by an institution. From a conceptual point of view, for many purposes the input perspective seems preferable over the output perspective. From a practical point of view, however, taking the input perspective often is not possible because the required data is not available. A more elaborate discussion of the pros and cons of productivity indicators and size-independent impact indicators can be found in a recent special section of *Journal of Informetrics* (Waltman, 2016b). In this special section, a discussion paper by Abramo and D'Angelo (2016a) argues in favor of the use of productivity indicators, while other contributions defend the use of size-independent impact indicators. We refer to Abramo and D'Angelo (2016b) for an institutional-level comparison between productivity indicators and size-independent impact indicators.

## 5. Evaluation of field-normalized indicators

The discussion in the previous section has shown that a large variety of field-normalized indicators have been proposed in the literature. This of course raises various questions: Do the indicators discussed in the previous section indeed provide properly field-normalized statistics? What are the advantages and disadvantages of the different ways in which field normalization can be performed? Is it possible to identify one specific approach to field normalization that can be considered superior over other approaches? To provide some partial answers to these questions, we now discuss the scientometric literature on the evaluation of field-normalized indicators. We restrict the discussion to indicators of impact.

To evaluate field-normalized indicators, some scientometricians choose to analyze the theoretical properties of indicators, while other scientometricians prefer to study the empirical characteristics of indicators. Different approaches to evaluate field-normalized indicators sometimes lead to different conclusions. For instance, from a theoretical perspective an indicator may seem appealing, while from an empirical perspective the same indicator may not seem very attractive. Below, we first discuss the theoretical evaluation of field-normalized indicators. We then turn to empirical evaluation.

**5.1. Theoretical evaluation of indicators**

In theoretical approaches to the evaluation of field-normalized indicators, the theoretical properties of indicators are studied. These are properties that do not depend on empirical data based on which indicators are calculated. After the theoretical properties of indicators have been established, the indicators are evaluated by deciding whether their properties are considered desirable or not. Whether a certain property is desirable is a subjective question that may legitimately be answered differently by different people. Theoretical evaluation therefore does not offer a universal and definitive answer to the question whether one indicator is superior over another. Instead, it aims to provide a deep understanding of the key differences between indicators. This may then guide users in choosing the indicator that best serves their needs.

In the calculation of the normalized citation score of a publication, defined as the ratio of the actual and the expected number of citations of a publication (see Subsection 4.1), theoretical considerations may help to choose between different ways in which the expected number of citations of a publication can be defined. The most common approach is to define a publication's expected citation count as the average citation count of all publications in the same field and in the same publication year.

An alternative approach is to consider in this definition only publications that have been cited at least once (Abramo et al., 2012a, 2012b). In the case of the former approach, for each combination of a field and a publication year, the average normalized citation score of all publications in that field and publication year equals exactly one. This may be regarded as an important property for a field-normalized indicator. The approach in which non-cited publications are left out from the definition of a publication's expected citation count does not have this property, which may be seen as a disadvantage of this approach.

Another issue in the calculation of the normalized citation score of a publication is the way in which publications belonging to multiple fields are handled. Based on the idea that the average normalized citation score of all publications in a field and publication year should equal one, it can be argued that the expected citation count of a publication belonging to multiple fields should be defined as the harmonic average of the expected citation counts corresponding to the different fields (Waltman et al., 2011a). However, a theoretical analysis presented by Smolinsky (2016) has shown that there are also other ways in which the expected citation count of a publication belonging to multiple fields can be defined. These alternative approaches lead to additional properties that may be considered attractive, but they have the disadvantage of introducing challenging computational issues.

As discussed in Subsection 4.1, there are different ways in which field-normalized indicators can be calculated at the aggregate level of for instance a research institution. The oeuvre argument of Moed (2010b) is a theoretical argument in favor of one approach, while the consistency argument of Waltman et al. (2011a) is a theoretical argument in favor of another approach. According to the oeuvre argument, it should not make a difference whether a citation is given to one publication in the oeuvre of a research unit or to some other publication in the same oeuvre. The basic idea of the consistency argument is that the ranking of two research units relative to each other should not change when both units make the same performance improvement. The oeuvre and consistency arguments can also be used to characterize some of the key differences between two versions of the SNIP journal impact indicator (Moed, 2010a; Waltman et al., 2013).

The choice of the counting method that is used to handle co-authored publications in the calculation of a field-normalized indicator can also be analyzed theoretically. When the full counting method is used, each publication is fully counted for each co-author, as explained in Subsection 4.1. On the other hand, when using a fractional counting method, co-authored publications are counted with a lower weight than publications that have not been co-authored. As pointed out by Waltman and Van Eck (2015), in the case of fractional counting, the mean normalized citation score indicator (see Subsection 4.1) has the property that the average value of the indicator for all research institutions active in a field equals exactly one. In the case of full counting, the indicator does not have this property. Using the full counting method, co-authored publications are counted multiple times, once for each of the co-authors. This double counting of co-authored publications, which tend to be publications that have received relatively large numbers of citations, has an inflanatory effect. It typically causes the mean normalized citation score indicator to have an average value for all research institutions active in a field that is above one. Because of this inflanatory effect, which is larger in some fields than in others, the full counting method provides statistics that are only partly field normalized. In order to obtain properly field-normalized statistics, a fractional counting method needs to be used. Alternatively, the use of a so-called multiplicative counting method (Perianes-Rodriguez & Ruiz-Castillo, 2015) can be considered.

**5.2. Empirical evaluation of indicators**

Empirical approaches to the evaluation of field-normalized impact indicators focus on three questions. First, assuming that a certain field classification system offers a satisfactory representation of scientific fields, which field-normalized indicators provide the best normalization? Second, to what extent do different field classification systems offer good representations of scientific fields, in particular for the purpose of field normalization? Third, which field-normalized indicators have the strongest correlation with peer review?

The idea of universality of citation distributions plays a key role in the literature dealing with the first question. Citation distributions are considered to be universal if the distribution of normalized citation scores is essentially identical for all scientific fields. The idea of universality of citation distributions has been introduced by Radicchi et al. (2008; see also Radicchi & Castellano, 2011). They claimed that universality of citation distributions can be achieved using a straightforward normalization approach in which the number of citations of each publication in a field is divided by the average number of citations of all publications in the field (excluding non-cited publications). However, in subsequent studies, it has been shown that this straightforward normalization approach yields citation distributions that are only approximately universal (Albarrán, Crespo, Ortuño, & Ruiz-Castillo, 2011; Bornmann & Daniel, 2009; Waltman, Van Eck, & Van Raan, 2012).

Based on the idea of universality of citation distributions, a so-called fairness test for field-normalized indicators has been proposed (Radicchi & Castellano, 2012a). This test has been used in various studies in which field-normalized indicators are compared (e.g., Leydesdorff, Radicchi, Bornmann, Castellano, & De Nooy, 2013; Radicchi & Castellano, 2012a, 2012c). The objective of having normalized citation distributions that are identical across fields also serves as the foundation of a methodology for quantifying the degree to which field-normalized indicators succeed in correcting for field differences (Crespo et al., 2013). This methodology has been applied in various studies (e.g., Crespo et al., 2013, 2014; Li, Radicchi, Castellano, & Ruiz-Castillo, 2013; Waltman & Van Eck, 2013b). Using the methodology of Crespo et al. (2013), it has been found that the normalization approach proposed by Radicchi and Castellano (2012c), based on a two-parameter power law transformation of citation counts, outperforms a number of other normalization approaches (Li et al., 2013). However, the standard approach of dividing the actual number of citations of a publication by the expected number of citations has also been shown to perform well. A study by Abramo et al. (2012a) in which a comparison is made of a number of field-normalized indicators is also based on the idea of trying to obtain normalized citation distributions that are identical across fields.

The above-mentioned studies assume that one has a satisfactory field classification system. They do not evaluate whether a certain classification system offers a good representation of scientific fields. This limited perspective on the evaluation of field-normalized indicators has been criticized by Sirtes (2012) in a letter commenting on Radicchi and Castellano (2012a; for a response, see Radicchi & Castellano, 2012b). According to Sirtes (2012; see also Waltman & Van Eck, 2013b), it is incorrect to evaluate a field-normalized indicator using the same classification system that is also used in the calculation of the indicator. This brings us to the second question raised in the beginning of this subsection: How suitable are different field classification systems for the purpose of field normalization?

Evaluations of the use of the WoS journal-based field classification system for the purpose of field normalization have been reported by Van Eck, Waltman, Van Raan,

Klautz, and Peul (2013) and Leydesdorff and Bornmann (2016). In both studies, the appropriateness of the fields in the WoS classification system for normalization purposes is questioned. For other studies questioning the use of the WoS classification system and proposing the use of alternative classification systems, we refer to Bornmann et al. (2008), Neuhaus and Daniel (2009), Van Leeuwen and Calero-Medina (2012), and Ruiz-Castillo and Waltman (2015). A systematic methodology for comparing the suitability of different classification systems for field normalization purposes has been presented by Li and Ruiz-Castillo (2013). We refer to Perianes-Rodriguez and Ruiz-Castillo (2017) for an application of this methodology.

Empirical approaches to the evaluation of field-normalized impact indicators also study the extent to which these indicators correlate with peer review. At the level of research programs and research departments in the natural sciences, indicators that use the standard normalization approach of dividing the actual number of citations of a publication by the expected number of citations have been shown to be moderately correlated with peer review assessments made by expert committees (Moed, 2005, Chapter 19; Rinia, Van Leeuwen, Van Vuren, & Van Raan, 1998). The correlation between normalized impact indicators and peer review has also been analyzed based on peer review outcomes from the Research Assessment Exercise (RAE) in the UK (Adams, Gurney, & Jackson, 2008). The main finding of this analysis is that impact indicators normalized at the level of journals hardly correlate with peer review, while impact indicators normalized at the level of journal-based fields in the WoS database or units of assessment in the RAE correlate significantly with peer review.

At the level of individual publications, the recently introduced relative citation ratio indicator has been claimed to be "well correlated" with expert judgments (Hutchins et al., 2016, p. 9). However, in a study by Bornmann and Haunschild (2017), the correlation between the relative citation ratio indicator and expert judgments has been characterized as "only low to medium" (p. 1064). In addition, it has been shown that, in terms of correlation with peer review, the relative citation ratio indicator has a performance that is similar to other field-normalized impact indicators. The studies by Hutchins et al. (2016) and Bornmann and Huanschild (2017) both make use of F1000 post-publication peer review data. Data from F1000 has also been used to analyze, at the level of individual publications, how strongly a number of field-normalized impact indicators correlate with peer review (Bornmann & Marx, 2015). It has been found that different field-normalized impact indicators all have a similar correlation with peer review. However, the authors leave open the possibility that F1000 data may not be sufficiently accurate to make fine-grained distinctions between different field-normalized impact indicators.

## 6. How much difference does it make in practice?

We have discussed a large number of field-normalized indicators as well as a large number of field classification systems that can be used by these indicators. We now consider the following question: How much difference does the choice of a field-normalized indicator, and possibly also a field classification system, make in practice, for instance when field-normalized indicators are used in the evaluation of research institutions, research groups, or individual researchers?

Various papers have presented analyses that provide insight into this question, most of them focusing on field-normalized indicators of impact. Before reporting our own analysis, we first briefly mention some of these papers, without going into the details of their findings. At the level of individual publications, the sensitivity of field-normalized indicators to the choice of a field classification system has been studied by Zitt et al. (2005). In the context of quantifying the impact of journals, field-normalized indicators that use a field classification system (see Subsections 4.1 and 4.2) have been compared with field-normalized indicators that use citing-side normalization (see Subsection 4.3) and that do not require a field classification system (Leydesdorff et al., 2013; Waltman & Van Eck, 2013a). Similar comparisons have also been made for indicators based on Mendeley bookmarks (Bornmann & Haunschild, 2016a). In the context of quantifying the impact of research institutions and their internal units, a number of studies have investigated for specific field-normalized indicators the effect of the choice of a field classification system. The use of the WoS journal-based classification system has been compared with the use of other less fine-grained journal-based systems (Colliander & Ahlgren, 2011; Glänzel et al., 2009), but also with the use of more fine-grained publication-based systems (Ruiz-Castillo & Waltman, 2015). In addition to analyzing the effect of the choice of a classification system, studies have also compared different normalization approaches for a given classification system. Perianes-Rodriguez and Ruiz-Castillo (2016) have for instance performed a comparison of two different ways in which normalized impact indicators can be obtained at the aggregate level of research institutions. Finally, as already mentioned, field-normalized indicators of productivity have received relatively limited attention in the literature. A comparison of two ways in which the FSS indicator (see Subsection 4.4) can be calculated at the level of research institutions has been reported by Abramo and D'Angelo (2015a).

**6.1. Empirical analysis of the sensitivity of field-normalized impact indicators to the choice of a field classification system**

Complementary to the studies mentioned above, we now present our own analysis. Our focus is on the sensitivity of field-normalized impact indicators to the choice of a field classification system. We are interested in particular in the sensitivity of the indicators at lower levels of aggregation, that is, at the level of internal units within a research institution. This level is highly relevant in practical applications of field-normalized impact indicators.

The mean normalized citation score, abbreviated as MNCS (Waltman et al., 2011a; see also Subsection 4.1), and the proportion of top 10% publications, abbreviated as PP(top 10%) (Waltman & Schreiber, 2013; see also Subsection 4.2), represent two of the most frequently used size-independent field-normalized impact indicators (taking into account also the use of variants of these indicators in commercial tools such as InCites and SciVal). Given the popularity of these indicators, it is important to understand their sensitivity to the choice of a field classification system. In this section, we therefore analyze the sensitivity of these indicators to the choice between on the one hand a traditional journal-based classification system, namely the classification system consisting of about 250 fields that is available in the WoS database, and on the other hand a publication-based classification system constructed algorithmically using the methodology of Waltman and Van Eck (2012). At the level of research institutions, the sensitivity to the choice between these two classification systems has been found to be relatively limited (Ruiz-Castillo & Waltman, 2015). However, this sensitivity has not yet been analyzed in a systematic way for smaller units. Below we present such an analysis for internal units within a large European university university.

Our analysis is based on the WoS database. More specifically, we use the Science Citation Index Expanded, the Social Sciences Citation Index, and the Arts & Humanities Citation Index. We use data on the publications of our focal university in the period 2010–2014. Publications are assigned to internal units within the university at three hierarchical levels. We refer to these levels as the faculty level, the department level, and the research group level. We take into account only units that have at least 50 publications. Also, only publications classified as research article or review article are considered. There are 13 faculties, 36 departments, and 130 research groups with 50 or more publications. Some basic statistics on the numbers of publications of these faculties, departments, and research groups are reported in Table 1.

Table 1. Statistics on the numbers of publications of the faculties, departments, and research groups of the focal university.

|  | Number of units | Number of publications | | | |
|---|---|---|---|---|---|
|  |  | Min. | Max. | Mean | Median |
| Faculties | 13 | 88 | 7626 | 1423.7 | 945.0 |
| Departments | 36 | 54 | 2785 | 560.3 | 322.5 |
| Research groups | 130 | 50 | 766 | 166.4 | 116.5 |

Citations are counted until the end of 2015. Author self-citations are excluded. In the case of the WoS journal-based classification system, publications in journals belonging to the *Multidisciplinary Sciences* category are reassigned to other categories based on their references. In the case of the publication-based classification system, we use a system that includes about 4000 fields. This is in line with the recommendation made by Ruiz-Castillo and Waltman (2015). The calculation of the MNCS and PP(top 10%) indicators is based on, respectively, Waltman et al. (2011a) and Waltman and Schreiber (2013). Normalization is performed for field and publication year, but not for publication type. A full counting approach is taken. Hence, each publication authored by a unit is fully counted for that unit, irrespective of possible co-authorship with other units inside or outside the focal university.

The results of the analysis are presented in Figure 1 for the 13 faculties, in Figure 2 for the 36 departments, and in Figure 3 for the 130 research groups. Each figure shows two scatter plots, one for the MNCS indicator and one for the PP(top 10%) indicator. In addition, Table 2 reports a number of statistics that summarize the differences between the results obtained using the WoS journal-based classification system and the results obtained using the publication-based classification system. For both the MNCS indicator and the PP(top 10%) indicator, the table presents the Pearson correlation between the results obtained using the two classification systems.

Moreover, the table also shows the mean absolute difference between the results, and the percentage of units for which the difference is considered to be large. A difference in the MNCS value of a unit of more than 0.5 is regarded as large. In the case of the PP(top 10%) indicator, we regard a difference of more than 5 percentage points as large.

Table 2. Statistics on the differences between the indicator values obtained using the WoS journal-based classification system and the indicator values obtained using a publication-based classification system.

|  | Pearson correlation | | Mean absolute difference | | % units with large difference | |
|---|---|---|---|---|---|---|
|  | MNCS | PP(top 10%) | MNCS | PP(top 10%) | MNCS | PP(top 10%) |
| Faculties | 0.89 | 0.80 | 0.12 | 2.0% | 0.0% | 0.0% |
| Departments | 0.95 | 0.93 | 0.12 | 1.8% | 2.8% | 5.6% |
| Research groups | 0.91 | 0.87 | 0.17 | 2.7% | 5.4% | 14.6% |

The results in Table 2 show that the mean absolute differences are larger at the level of the research groups than at the level of the faculties and the departments. Likewise, the percentage of units with large differences is highest at the research group level. These findings may not be surprising. Research groups on average have a much smaller number of publications than faculties and departments (see Table 1), and therefore the MNCS and PP(top 10%) values of research groups can be expected to be more sensitive to the choice of a classification system than the corresponding values of faculties and departments. Based on the results in Table 2 and in Figures 1, 2, and 3, it can also be concluded that the PP(top 10%) indicator is more sensitive to the choice of a classification than the MNCS indicator.

Based on our results, how sensitive are field-normalized impact indicators to the choice of a field classification system? The answer to this question may depend on the expectations that one has. Some readers may consider the differences between the results obtained using the WoS journal-based classification system and the results obtained using the publication-based classification system to be within an acceptable margin. Others may be concerned to see, for instance, that for about one out of seven research groups the PP(top 10%) indicator increases or decreases by more than 5 percentage point when changing the classification system based on which the indicator is calculated (see Table 2). Our perspective is that the results illustrate the risk of overinterpreting field-normalized indicators, especially at lower levels of aggregation, such as the research group level. There is no perfect way to correct for differences between fields. Different field normalization approaches make different choices in how they correct for field differences. Each approach is informative in its own way. When working with one specific field normalization approach, it is essential to keep in mind that this approach offers just one perspective on field normalization and that other approaches will give a different perspective, in some cases even a perspective that may be different in a quite fundamental way.

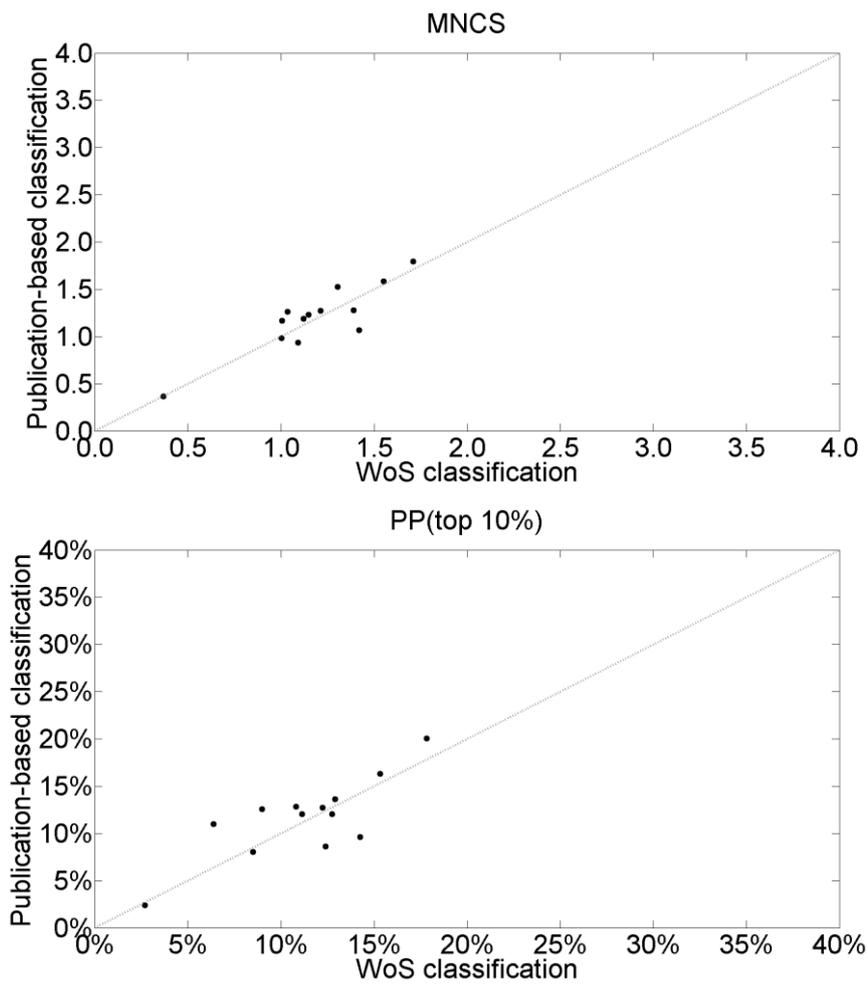

Figure 1. Scatter plots of the MNCS and PP(top 10%) values of the 13 faculties of the focal university, obtained using either the WoS journal-based classification system or a publication-based classification system.

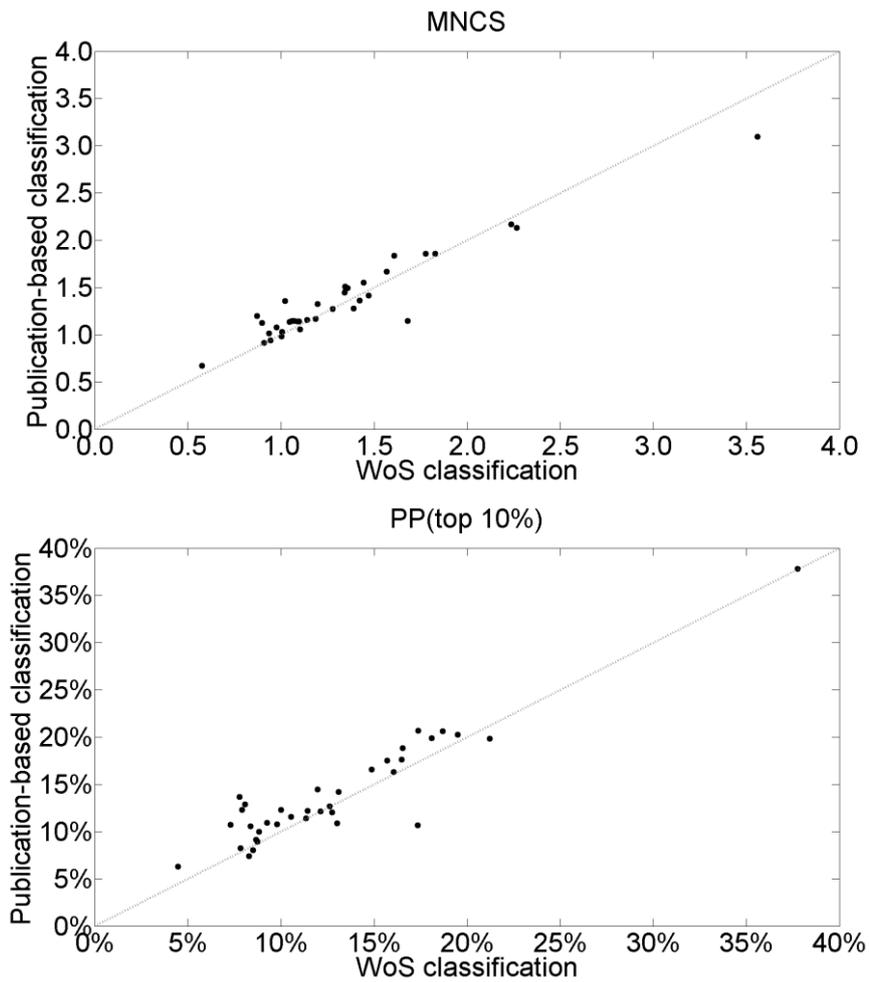

Figure 2. Scatter plots of the MNCS and PP(top 10%) values of the 36 departments of the focal university, obtained using either the WoS journal-based classification system or a publication-based classification system.

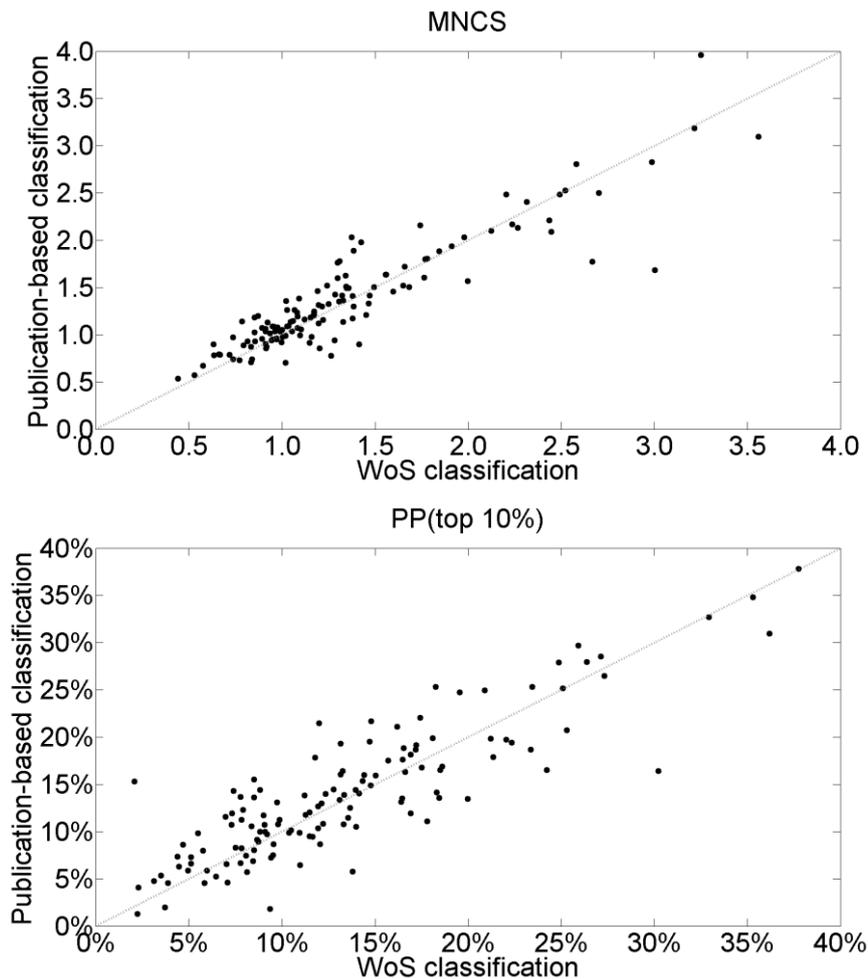

Figure 3. Scatter plots of the MNCS and PP(top 10%) values of the 130 research groups of the focal university, obtained using either the WoS journal-based classification system or a publication-based classification system.

## 7. Conclusion

Some critics question whether field normalization is truly attainable. In the literature, this viewpoint is represented by Kostoff (2002) and Kostoff and Martinez (2005), who criticize the idea of field-normalized impact indicators, arguing that citation counts of publications should be compared only if publications are very similar to each other. According to Kostoff and Martinez (2005, p. 61), "a meaningful 'discipline' citation average may not exist, and the mainstream large-scale mass production semi-automated citation analysis comparisons may provide questionable results". In principle, critics make a valid point. Taking their position to the extreme, one could argue that every publication is unique in its own way and, consequently, that any comparison of citation counts of publications is problematic. Likewise, it could be argued that every researcher is unique and that any comparison of publication and citation counts of researchers is therefore in some sense unfair.

However, one may also take a more pragmatic perspective on the idea of field normalization. In managing and evaluating scientific research, there often is a need to compare different research units (e.g., research institutions, research groups, or individual researchers). Scientometric indicators of course provide an incomplete picture of the units to be compared. Moreover, these indicators are affected by all kinds of distorting factors, for instance related to the characteristics of the underlying data sources, the peculiarities of the units to be compared, and the nature of the scientific fields in which these units are active. Nevertheless, despite their limitations, scientometric indicators provide useful and relevant information for supporting the management and evaluation of scientific research. In many cases, the usefulness of scientometric indicators can be increased by making corrections for some of the most significant distorting factors, and field differences typically are one such a factor. Field normalization does not correct for all distorting factors, but it corrects at least partly for one of the most important ones. From this point of view, field normalization serves an important practical purpose.

**7.1. Strengths and weaknesses of different field normalization approaches**

We have provided an overview of a large number of approaches to field normalization. Although some field normalization approaches can be considered superior over others, we do not believe there to be a single optimal approach. Instead, there is a trade-off between the strengths and weaknesses of different approaches. Some field normalization approaches have a high level of technical sophistication. These approaches may for instance use an algorithmically constructed publication-based field classification system, or they may not need a classification system at all, and instead of the traditional full counting method these approaches may use a fractional counting method for dealing with co-authored publications. Other field normalization approaches are much more basic. They for instance rely on the standard journal-based classification system made available in a database such as WoS or Scopus, and they handle co-authored publications using the standard full counting method. In general, the more sophisticated approaches can be expected to better correct for field differences than the more basic approaches. On the other hand, however, the more basic approaches tend to be easier to understand and more transparent. This enables users to carefully reflect on what a field-normalized indicator does and does not tell them, and it allows users to recognize the limitations of the indicator. The more sophisticated approaches tend to be black boxes for many users, forcing users to blindly trust the outcomes provided by these approaches. Due to the low level of transparency, it is difficult for users to understand the limitations of the more sophisticated approaches and to interpret the outcomes obtained using these approaches in the light of these limitations.

As a general rule, in situations in which in-depth reflection on scientometric indicators is desirable or even essential, for instance when indicators are used to support the evaluation of individual researchers, we recommend the use of simple and transparent field normalization approaches. Complex non-transparent approaches should not be used in such situations. On the other hand, there are also situations in which the use of more advanced field normalization approaches, possibly with a relatively low level of transparency, may be preferable. This could be the case in situations in which scientometric indicators are used at a high level of aggregation, for instance at the level of entire research institutions or countries, where in-depth reflection on the indicators may hardly be possible, or in situations in which scientometric indicators are used in a purely algorithmic way, for instance when they are embedded in a funding allocation model.

**7.2. Contextualization as an alternative way to deal with field differences**

We end this chapter by pointing out that field normalization is not the only way to deal with field differences in scientometric analyses. When detailed assessments need to be made at the level of individual researchers or research groups, an alternative approach is to use straightforward non-normalized indicators and to contextualize these indicators with additional information that enables evaluators to take into account the effect of field differences (see also Waltman & Van Eck, 2016). For instance, to compare the productivity of researchers working in different fields, one could present non-normalized productivity indicators (e.g., total publication or citation counts during a certain time period) for each of the researchers to be compared. One could then contextualize these indicators by selecting for each researcher a number of relevant peers working in the same field and by presenting the productivity indicators also for these peers. In this way, each researcher's productivity can be assessed in the context of the productivity of a number of colleagues that have a reasonably similar scientific profile.

An advantage of the above contextualization approach could be that it may lead to a less mechanistic way of dealing with field differences. In our experience, field-normalized indicators tend to be used quite mechanistically, with little attention for their limitations. This is problematic especially at lower levels of aggregation, for instance at the level of individual researchers or research groups, where field-normalized indicators are quite sensitive to methodological choices, such as the choice of a field classification system. In a research evaluation, the contextualization approach outlined above may encourage evaluators to reflect more deeply on the effect of field differences and to perform inter-field comparisons in a more cautious and thoughtful way. It may also invite evaluators to combine scientometric evidence of field differences with their own expert knowledge of publication, collaboration, and citation practices in different fields of science. Hence, peer review and scientometrics may be used together in a more integrated manner, which can be expected to improve the way in which research is evaluated (Waltman & Van Eck, 2016).

## Acknowledgment

We would like to thank Lutz Bornmann, Robin Haunschild, Loet Leydesdorff, Javier Ruiz-Castillo, and an anonymous referee for their helpful comments on an earlier version of this chapter. These comments have led to numerous improvements.

Vaccario, G., Medo, M., Wider, N., & Mariani, M.S. (2017). Quantifying and suppressing ranking bias in a large citation network. *Journal of Informetrics*, *11*(3), 766–782.

Van Eck, N.J., Waltman, L., Van Raan, A.F.J., Klautz, R.J.M., & Peul, W.C. (2013). Citation analysis may severely underestimate the impact of clinical research as compared to basic research. *PLOS ONE*, *8*(4), e62395.

Van Leeuwen, T.N., & Calero Medina, C. (2012). Redefining the field of economics: Improving field normalization for the application of bibliometric techniques in the field of economics. *Research Evaluation*, *21*(1), 61–70.

Van Leeuwen, T.N., Visser, M.S., Moed, H.F., Nederhof, T.J., & Van Raan, A.F.J. (2003). The Holy Grail of science policy: Exploring and combining bibliometric tools in search of scientific excellence. *Scientometrics*, *57*(2), 257–280.

Van Raan, A.F.J. (2005). Measuring science: Capita selecta of current main issues. In H.F. Moed, W. Glänzel, & U. Schmoch (Eds.), *Handbook of quantitative science and technology research* (pp. 19–50). Springer.

Van Raan, A.F.J., Van Leeuwen, T.N., Visser, M.S., Van Eck, N.J., & Waltman, L. (2010). Rivals for the crown: Reply to Opthof and Leydesdorff. *Journal of Informetrics*, *4*(3), 431–435.

Vîiu, G.-A. (2017). Disaggregated research evaluation through median-based characteristic scores and scales: A comparison with the mean-based approach. *Journal of Informetrics*, *11*(3), 748–765.

Vinkler, P. (2012). The case of scientometricians with the "absolute relative" impact indicator. *Journal of Informetrics*, *6*(2), 254–264.

Waltman, L. (2016a). A review of the literature on citation impact indicators. *Journal of Informetrics*, *10*(2), 365–391.

Waltman, L. (2016b). Special section on size-independent indicators in citation analysis. *Journal of Informetrics*, *10*(2), 645.

Waltman, L., & Schreiber, M. (2013). On the calculation of percentile-based bibliometric indicators. *Journal of the American Society for Information Science and Technology*, *64*(2), 372–379.

Waltman, L., & Van Eck, N.J. (2010). The relation between eigenfactor, audience factor, and influence weight. *Journal of the American Society for Information Science and Technology*, *61*(7), 1476–1486.